\documentclass[twocolumn]{aastex631}

\begin{document}
	
	\title{ On the Interacting/Active Lifetime of Supernova Fallback Disk around Isolated Neutron Stars }
	
	\author[0000-0002-9739-8929]{Kun Xu}
	\affiliation{School of Science, Qingdao University of Technology, Qingdao 266525, China; xukun@smail.nju.edu.cn}
	
	\author[0000-0001-5532-4465]{Hao-Ran Yang}
	\affiliation{Department of Astronomy, Nanjing University, Nanjing 210023, People’s Republic of China}
	\affiliation{Key Laboratory of Modern Astronomy and Astrophysics, Nanjing University,
		Ministry of Education, Nanjing 210023, People’s Republic of China}
	
	\author[0000-0002-2479-1295]{Long Jiang}
	\affiliation{School of Science, Qingdao University of Technology, Qingdao 266525, China}
	\affiliation{School of Physics and Electrical Information, Shangqiu Normal University, Shangqiu 476000, China}
	
	\author[0000-0002-0785-5349]{Wen-Cong Chen}
	\affiliation{School of Science, Qingdao University of Technology, Qingdao 266525, China}
	\affiliation{School of Physics and Electrical Information, Shangqiu Normal University, Shangqiu 476000, China}
	
	\author[0000-0002-0584-8145]{Xiang-Dong Li}
	\affiliation{Department of Astronomy, Nanjing University, Nanjing 210023, People’s Republic of China}
	\affiliation{Key Laboratory of Modern Astronomy and Astrophysics, Nanjing University,
		Ministry of Education, Nanjing 210023, People’s Republic of China}
	
	\author{Jifeng Liu}
	\affiliation{School of Astronomy and Space Sciences, University of Chinese Academy of Sciences, Beijing, People’s Republic of China}
	\affiliation{Key Laboratory of Optical Astronomy, National Astronomical Observatories, Chinese Academy of Sciences, Beijing, People’s Republic of China}
	\affiliation{WHU-NAOC Joint Center for Astronomy, Wuhan University, Wuhan, People’s Republic of China}
	
	\begin{abstract}
		
		The fallback disk model is widely accepted to explain long-period neutron stars (NSs) which can’t be simulated by magnetic dipole radiation.
		However, no confirmed detection of disk was found from the newly discovered long period pulsars GLEAM-X 162759.5-523504.3, GPM J1839-10 and the known slowest isolated NSs 1E 161348-5055.
		This might be that the disks have either been in noninteracting/inactive state where its emission is too weak to be detected or have been disrupted.
		In this work, we conduct simulations to examine the lifetime of supernova fallback disks around isolated neutron stars.
		We assume that the disk's mass varies in a self-similar way and its interaction with the NS occurs only in interacting/active state.
		Our results reveal that nearly all the interacting lifetimes for the disk are shorter than $10^5$ yr while the existence lifetimes are considerably longer.
		
	\end{abstract}
	

	\section{Introduction} \label{sec:intro}

	The fallback disk model is widely employed to explain several unique observations of neutron stars (NSs), like the spin evolution of certain long period NSs \citep{DeLuca2006,Xu2019}, the peak following the main one in SN light curve \citep{Lin2021} and so on. 
	In addition, a fallback disk is also the presumed origin of planets around an NS
	\citep[e.g.][]{Wolszczan1992,Wolszczan1994,Nitu2022,Mishra2023}.
	A disk can result from the supernova (SN) explosion if the newborn NS captured part of the ejected material due to its gravity, provided it possesses sufficient angular momentum \citep[e.g.][]{Woosley1993,Alpar2001,Benli2016}.
	Then it will interact with the NS in accretion or propeller phase, potentially emitting radiation in X-ray band.
	Although there is no interaction between the disk and the NS in ejector phase or dead state, the disk may still be detected in infrared (IR) band if it was heated by the X-ray from the NS, which may be powered by the magnetic energy.
	
	A fallback disk was first reported to be discovered around an anomalous X-ray pulsar (AXP) 4U 0142+61 (hereafter 4U 0142) in mid-infrared \citep{Wang2006}, which is a cool disk with estimated mass $\sim 10^{-5} {\rm ~M_{\odot}}$ and lifetime $\gtrsim 10^{6} {\rm ~yr}$.
	Up to now, in addition to 4U 0142, IR counterparts have been confidently detected in 1RXS J170849.0-400910 \citep{Durant2006}, XTE J1810-197 \citep{Testa2008}, 1E 2259+586 \citep{Kaplan2009}, Vela and Geminga pulsars \citep{Danilenko2011}, etc, indicating that fallback disks may be present in these sources.
	
	Fallback disk around magnetar is the most popular model to explain long period NSs which can't be simulated by magnetic dipole radiation.
	However, despite numerous deep IR observations, the existence of a fallback disk around the slowest known isolated pulsar 2E1613.5-5053 (also known as 1E 161348-5055, hereafter 1E 1613) with spin period $P_{\rm spin} \approx 6.67 {\rm ~h}$ remains unconfirmed \citep{Tendulkar2017}.
	1E 1613 locates in the center of a young supernova remnant RCW 103 with age less than a few thousand years \citep{DeLuca2006}, which is believed to be spun down by a fallback disk \citep[e.g.][]{DeLuca2006,Li2007,Tong2016,Ho2017,Xu2019}.
	Recently, two additional long period radio pulsars, GLEAM-X 162759.5-523504.3 (hereafter J1627) with $P_{\rm spin} \approx 18 {\rm ~min}$ \citep{Hurley-Walker2022} and GPM J1839-10 (hereafter J1839) with $P_{\rm spin} \approx 21 {\rm ~min}$ \citep{Hurley-Walker2023}, were reported.
	Their spin periods exceed the upper limit for NSs governed by magnetic dipole radiation ($\sim 248 {\rm ~s}$ \footnote{It's an analytical solution for the spin evolution of an NS assuming an unchanged magnetic field of $10^{16} {\rm ~G}$ within $10 {\rm ~kyr}$ \citep{Xu2021}, and the simulation results in this work ($\sim 245 {\rm ~s}$) align with it when the magnetic field decay is taken into account in a long timescale.}), while no observational evidence of fallback disk has been detected for them.
	
	Why we can't detect the fallback disk around long period NSs?
	We think the most likely cause is the disk may have either been in noninteracting/inactive state where its emission is too weak to be detected or has been disrupted.
	In order to validate this idea, we conduct simulations of the lifetime of fallback disks in this work. 
	
	This paper is organized as follows. We set up the fallback disk model in Section 2 and exhibit the results in Section 3. The summary and discussion are in Section 4.
	
	\section{The Model}
	
	In our model, the physical parameters of the fallback disk are considered varying in a self-similar way, including the disk mass $M_{\rm d}$, the mass loss rate of the disk $\dot{M}$ and the outer radius of the disk $R_{\rm out}$, 
	\begin{equation}
		\begin{array}{l}
			M_{\rm d}(t) = M_{\rm d,0} (t/t_{\rm f})^{-a+1}, \\
			\dot{M}(t) = \dot{M_0} (t/t_{\rm f})^{-a}, \\
			R_{\rm out}(t) = R_{\rm f} (t/t_{\rm f})^{2a-2},
		\end{array}
	\end{equation}
	here $M_{\rm d,0}$ and $R_{\rm f}$ are the initial values of $M_{\rm d}$ and $R_{\rm out}$.
	$t_{\rm f}$ is the disk formation time, which is the maximum value of the dynamical and viscous timescale \citep{Xu2019}, 
	and $\dot{M}_0 = M_{\rm d,0}/t_{\rm f}$ is the initial mass loss rate.
	The power law index of mass loss rate $a$ is thought to be relative to the opacity and wind loss of the disk \citep[e.g.][]{Cannizzo1990,Li2007,Tong2016,Lin2021}.
	We assume that the disk wind loss rate is $(1-\eta)\dot{M}(t)$, so the mass flow rate in the disk is $\eta \dot{M}(t)$ and we assume $\eta$ as a constant for simplicity.
	
	Based on the model in \citet{Xu2019}, we add some limitations on the evolution to simulate the lifetime of the interacting fallback disk.
	The disk is thought to be formed at time $t_{\rm f}$, then starts to interact with the NS when it enters interacting state and stops when it leaves interacting state (i.e., it enters the noninteracting state).
	In order to describe the disk state, we introduce three important radii,
	\begin{enumerate}
		\item The inner radius of the disk
		\begin{equation}
			R_{\rm in} = \xi (\frac{\mu^4}{2 G M_{\rm NS} \dot{M}_{\rm in}^2})^{1/7}, 
		\end{equation}
		where
		$\mu = B R_{\rm NS}^3$ is the magnetic dipole moment of the NS.
		$M_{\rm NS} = 1.4 M_{\odot}$ and $R_{\rm NS}=10^6 {\rm ~cm}$ are the typical mass and radius of an NS, respectively.
		$G$ is the gravitational constant and $\dot{M}_{\rm in}$ is mass flow rate at the inner radius of the disk, which is thought to be equal to $\eta \dot{M}(t)$ in our model.
		$\xi$ is the correction parameter between the inner radius and the traditional Alfv\'en radius for spherical accretion $R_{\rm A} = (\frac{\mu^4}{2 G M_{\rm NS} \dot{M}_{\rm in}^2})^{1/7}$. 
		\item The corotation radius
		\begin{equation}
			R_{\rm c} = (\frac{GM_{\rm NS}}{\Omega_{\rm s}^2})^{1/3}, 
		\end{equation}
		where the angular velocity $\Omega_{\rm s}$ of the NS is equivalent to the Keplerian angular velocity $\Omega_{\rm K}(R)$ of the disk.
		\item The light cylinder radius
		\begin{equation}
			R_{\rm lc} = \frac{c}{\Omega_s}, 
		\end{equation}
		where the corotation velocity equals the speed of light $c$.
	\end{enumerate}
	Then we can use them to define the disk phases.
	Accretion occurs when the inner radius of the disk is smaller than the corotation radius in the accretion phase, while the material are tossed out of the system in the propeller phase where the inner radius larger than the corotation radius and smaller than the light cylinder radius. When the inner radius larger than the light cylinder radius, no material in the disk interacts with the NS in the ejector phase.
	In brief,  the disk phases can be described by the following equations
	\begin{equation}
		\left\{ 
		\begin{array}{l}
			R_{\rm in}<R_{\rm co} {\rm , ~~~~~~~~~accretion ~phase}; \\
			R_{\rm co}<R_{\rm in}<R_{\rm lc} {\rm , ~propeller ~phase};  \\
			R_{\rm in}>R_{\rm lc} {\rm , ~~~~~~~~~~ejector ~phase}. 
		\end{array}
		\right.
	\end{equation}
	The disk in interacting state represents it has interplay with the NS, i.e., the disk in the accretion phase or the propeller phase. 
	The torque exerted on the NS in interacting state is
	\begin{equation}
		N = \pm \dot{M}_{\rm in} (G M_{\rm NS} R_{\rm in})^{1/2} ,
	\end{equation}
	where the positive value is for the accretion phase and the negative one, for simplicity, is for the propeller phase \footnote{Our simplified propeller form does not differ significantly from others, e.g., $N_{\rm prop} = 2 \dot{M} R_{\rm in}^2 (\Omega_{\rm K}(R_{\rm in}) - \Omega)$ in \cite{2024ApJ...967...24F}. } \citep{Xu2021}.

	The conditions for interacting/active disk are:
	\begin{enumerate}
		\item The disk can form, i.e., the initial inner radius derived from the input parameters is smaller than the initial outer radius $R_{\rm f}$.
		\item The formation time $t_{\rm f}$ is smaller than the universe lifetime, which is set to be 10 Gyr in our simulation.
		\item The inner radius $R_{\rm in}$ is smaller than the interacting outer radius $R_{\rm out,interact} = \min(R_{\rm out,act}, R_{\rm lc})$ during the evolution, where $R_{\rm out,act} = \min(R_{\rm out}, R_{\rm sg}, R_{\rm neu})$ \citep{Xu2019}, $R_{\rm sg}$ \citep{Burderi1998} and $R_{\rm neu}$\citep{Liu2015}  are the active outer radius, the self-gravity radius and the neutralization radius
		\footnote{The disk becomes neutralization when its temperature lower than a critical value $T_{\rm C} \sim 6500 {\rm K}$ \citep{Menou2001}. 
			However, it has been suggested that the effects of X-ray irradiation of the disk by the accretion and cooling luminosities may lead the inactive disk to appear at a lower temperature \citep{Alpar2001,Inutsuka2005,Alpar2013}. 
			So we take $T_{\rm C} \sim 300 {\rm ~K}$ \citep{Inutsuka2005} in our simulation.}, 
		respectively.
		\item The disk mass is larger than $10^{-10} {\rm ~M_{\odot}}$ \footnote{We think the disk will disintegrate if its mass is smaller than $10^{-10} {\rm ~M_{\odot}}$.}.
	\end{enumerate}
	In contrast, there is no effect on the NS from the disk in the noninteracting/inactive state, which includes the ejector phase and the dead state (the state when $R_{\rm in}>R_{\rm out,act}$).
	So the torque exerted on the NS in noninteracting state is the magnetic dipole radiation one
	\begin{equation}
		N_{\rm B} = - \frac{2 \mu^2 \Omega_{\rm s}^3}{3 c^3}.
	\end{equation}

	We can observe the X-ray radiation resulting from the interaction between the disk and the NS, while the noninteracting disk dose not power the NS's X-ray emission. However, we still have opportunities to detect it in the infrared, as demonstrated by the debris disk surrounding 4U 0142+61 \citep{Wang2006}.
	We calculate two lifetime for the fallback disk, which are $\tau_{isd}$ for disk in interacting state and $\tau_{exd}$ for existence time of disk from formed to disappeared.
	
	The magnetic field decay of an NS is considered in the model since its time is setting to be as long as 10 Gyr, which follows $B(t) = B_{\rm i} (1+ \alpha t/\tau_{\rm d,i})^{-1/\alpha}$ \citep{Dall'Osso2012,Xu2023}, where $B_{\rm i}$ is the initial magnetic field when the NS born and $\tau_{\rm d,i} = \tau_{\rm d} /(B_{\rm i}/10^{15} {\rm ~G})^{\alpha}$ \citep{Fu2012} and $\tau_{\rm d}$ is the field decay timescale.
	Considering that the corotation radius $R_{\rm c}$ and the light cylinder radius $R_{\rm lc}$ are relative to the spin period of an NS, the spin evolution is also simulated.

	Given the uncertainties about which parameters are dominant, we introduce 9 input parameters in our model, which are 
	the initial spin period of an NS $P_{\rm 0}$,
	the initial magnetic field of an NS $B_{\rm i}$,
	the characteristic decay timescale of NS's magnetic field $\tau_{\rm d}$,
	the power law index parameter of magnetic field decay $\alpha_{\rm B}$,
	the correction parameter between the inner radius and the traditional Alfv\'en radius for spherical accretion $\xi$,
	the initial disk mass $M_{\rm d,0}$, 
	the initial outer radius of the disk $R_{\rm f}$ \footnote{ $R_{\rm f}$ in units of $R_{\rm S} \approx 4\times 10^{5} {\rm ~cm}$, which is the Schwartzschild radius for a $1.4{\rm ~M_{\odot}}$ NS and the lower limit of $R_{\rm f}$ is the characteristic radius of an NS $R_{\rm NS}$.},
	the fraction between the mass flow rate in the disk and the mass loss rate of the disk $\eta$.
	and the power law index of mass loss rate of the disk $a$,
	Their ranges are showed in Table \ref{table:param}.

\begin{deluxetable}{cc}
		\tablewidth{0pt}
		\tablecaption{Input parameters \label{table:param}} 
		\tablehead{\colhead{name} & \colhead{range} }
	    \startdata
		$P_{\rm 0}$ (${\rm s}$)   			 & $[10^{-3}, 1]$  \\
		$B_{\rm i}$ (${\rm G}$)			 & $[10^{11}, 10^{16}]$ \\
		$\tau_{\rm d}$ (${\rm yr}$) 		 & $[10, 10^5]$  \\
		$\alpha_{\rm B}$							 & $[1.0, 1.8]$ \\
		$\xi$                                & $[0.1,1.0]$ \\
		$M_{\rm d,0}$ (${\rm M_{\odot}}$)	 & $[10^{-10}, 10^{-1}]$  \\
		$R_{\rm f}$							 & $[{R_{\rm NS}}, 10^6 {R_{\rm S}} ]$ \\
		$\eta$						 		 & $[0.1, 1]$ \\
		$a$                                  & $[19/16,7/3]$ \\
		\hline
	    \enddata
\end{deluxetable}

	\section{Results}
	
	\begin{figure}
		\plotone{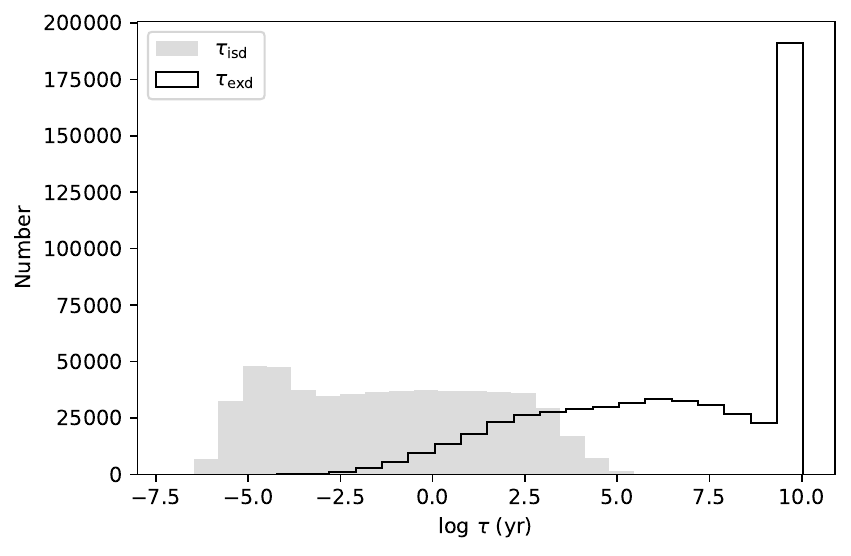}
		\plotone{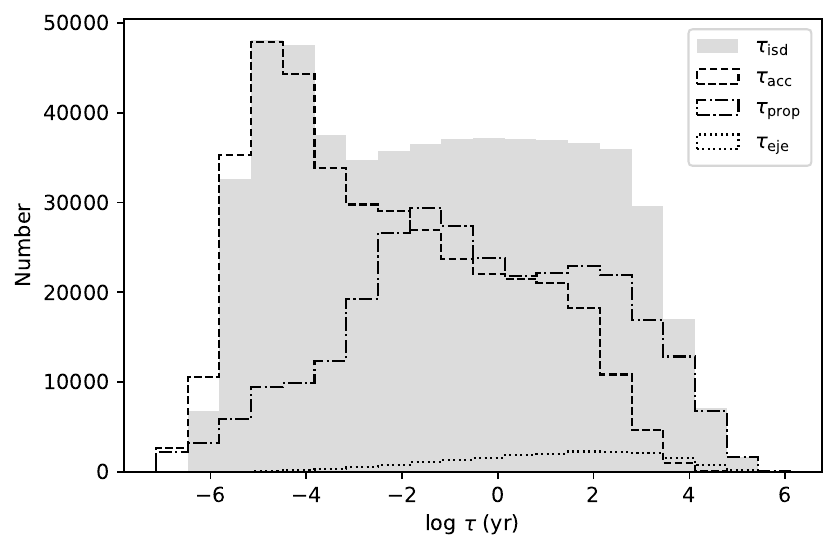}
		\caption{The histograms of the fallback disk lifetime. The gray bar-type histogram indicates the interacting lifetime $\tau_{isd}$ distribution in both panels. 
			The step-type histograms illuminate the existence lifetime $\tau_{exd}$ (the solid line) distribution in the upper panel and the phase lifetime distribution of interacting disks in the lower panel, including the lifetime for the accretion phase $\tau_{\rm acc}$ (the dashed line), the propeller phase $\tau_{\rm prop}$ (the dotted-dashed) and the ejector phase $\tau_{\rm eje}$ (the dotted line).
			\label{fig:hist}}
	\end{figure}
	
	We evolve a sample of $10^6$ typical NSs with mass $M_{\rm NS} = 1.4 {\rm ~M_{\odot}}$ and radius $R_{\rm NS} = 10^6 {\rm ~cm}$.
	The evolution starts from the NS born and stops when the disk is disrupted or the system lifetime reaches the universe lifetime.
	We exhibit the histograms for fallback disk lifetime in Figure \ref{fig:hist}.
	In the upper panel, the gray bar-type histogram indicates the distribution of the interacting lifetime $\tau_{\rm isd}$ as well as in the lower panel, and the black solid step-type histogram shows the lifetime of the disk's existence $\tau_{\rm exd}$.
	The lower panel displays the lifetime of the interacting disk in different phases, with step-type histograms showing the lifetime for the accretion phase $\tau_{\rm acc}$, propeller phase $\tau_{\rm prop}$, and ejector phase $\tau_{\rm eje}$, denoted by dashed, dotted-dashed and dotted lines, respectively.
	It reveals that 
	\begin{enumerate}
		\item Nearly all the interacting lifetimes are shorter than $10^5 {\rm ~yr}$ for the disk and more than half of them are shorter than $1 {\rm ~yr}$. The maximum interacting lifetime is about $3.5 {\rm ~Myr}$.
		\item The existence lifetimes are considerably longer than their interacting lifetimes, with over half exceeding $10^5 {\rm ~yr}$ and nearly one third of them reaching the universe lifetime.
		\item The sources experienced ejector phases are significantly less than those undergoing accretion or propeller phases.
		\item The lifetimes of disks in accretion phase are much shorter than those in propeller or ejector phase. Less than a quarter of accretion lifetimes are longer than $1 {\rm ~yr}$, while nearly half of propeller and three quarters of ejector lifetimes exceed this age. The maximum lifetime of accretion phase, propeller phase and ejector phase are about $0.6$, $ 3.5 $ and $ 1.8 {\rm ~Myr}$, repsctively.
	\end{enumerate}

	\section{Discussion}
	
	\begin{figure}
		\plotone{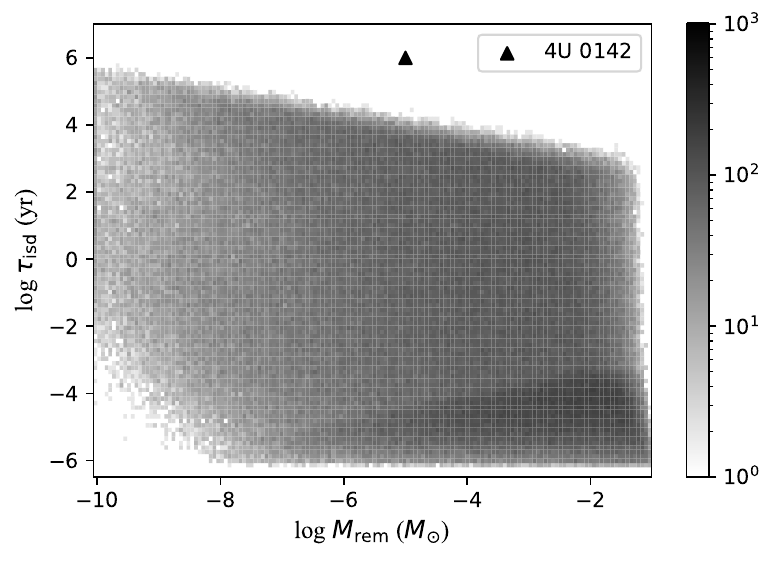}
		\caption{The 2D histogram for the interacting lifetime $\tau_{\rm isd}$ against the remnant mass $M_{\rm rem}$ when the fallback disk leaving interacting state. The triangle indicates the estimated mass and lifetime of 4U 0142.
			\label{fig:tm_4u0142}}
	\end{figure}
	
	A fallback disk around an NS is very hard to be detected, let alone the detection of its properties.
	So far, 4U 0142 is the unique source with estimated lifetime and mass of its fallback disk.
	Observations reveal that it is a cool disk in the infrared and does not contribute to the X-ray emission of the NS \citep{Wang2006}, i.e., the disk is very likely in noninteracting state.
	We display the interacting lifetime $\tau_{\rm isd}$ against the remnant mass $M_{\rm rem}$ when the fallback disk leaving interacting state in Figure \ref{fig:tm_4u0142}, where the triangle indicates the estimated values of 4U 0142 \citep{Wang2006}.
	It shows that the triangle is out of the histogram region, confirming that the disk around 4U 0142 is not in the interacting state.
	This is corresponding to the observations.
	However, the disk around 4U 0142+61 was also modeled from optical to mid-IR by a viscously active and irradiated disk spectrum \citep{Ertan2007} while the accretion is the source of the X-ray luminosity. The phase-dependent pulsed X-ray spectrum of this source was also shown to be in agreement with the emission from the accretion column and the NS surface \citep{Trumper2010}. The disk powering the pulsar's X-ray emission indicates that the disk is in active state in our model, which also supports the existence of the disk.

	\begin{figure}
		\plotone{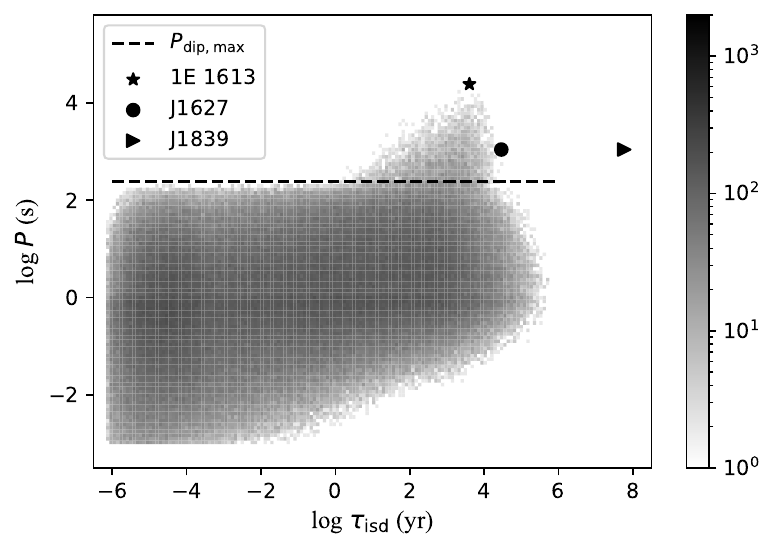}
		\caption{The 2D histogram for neutron stars' spin period $P$ against the interacting lifetime of fallback disks $\tau_{\rm isd}$. The dashed line indicates the maximum spin period of NSs without disk $P_{\rm mnd}$.
			\label{fig:p_tau}}
	\end{figure}
	
	Figure \ref{fig:p_tau} exhibits the spin periods of NSs against the interacting lifetime of fallback disks. 
	The dashed line represents the spin period of NSs without disks, $P_{\rm dip, max} \approx 245 {\rm ~s}$\footnote{This is consistent with the estimated value from analytical solutions ($\sim 248 {\rm ~s}$) in \citep{Xu2021}. }.
	NSs' spin periods can exceed this value when the interacting lifetime of fallback disks $\tau_{\rm isd}$ is in the range of  $1 {\rm ~yr}$  to a few $\times 10 {\rm ~kyr}$ with magnetic fields $B \geq 10^{14} {\rm ~G}$ and can reach $10^4 {\rm ~s}$ when $\tau_{\rm isd} \sim 10 {\rm ~kyr}$.
	The maximum $P$ in the simulation is $\approx 2.8 \times 10^4 {\rm ~s} \approx 7.8 {\rm ~hr}$ with $ \tau_{\rm isd} \approx 6.2 {\rm ~kyr}$.
	The three special NSs with spin periods longer than $P_{\rm dip, max}$ are indicated by star for 1E 1613, circle for J1627 and triangle for J1839\footnote{The age of 1E 1613 in Figure \ref{fig:p_tau} is taken to be $4000 {\rm ~yr}$ which is the age of its associated supernova remnant (SNR) RCW 103, while the age of J1627 and J1839 is their spin-down age since no associated SNRs found.}, respectively.
	It shows that J1839 significantly departs from the histogram region, suggesting that either its disk is no longer in active state or it has no disk. This aligns with the observation that J1839 is a radio pulsar without detected X-ray emissions. Given its spin period surpassing $P_{\rm dip, max}$, it is plausible that J1839 has had an active disk spinning it down to a very long period and then the disk evolved to noninteracting state or even been disrupted.
	1E 1613 and J1627 reside near the edge of the histogram region, suggesting that the fallback disks around them may be in the propeller phase with X-ray emission too weak to be detected in observation or in the noninteracting state without X-ray emission. 
	J1627 exemplary exhibits the latter behavior as a radio pulsar without X-ray detection, while 1E 1613 is a X-ray source with emission believed to originate from magnetic energy release rather than from the disk. This supports the results of our simulation.
	Above all, the extremely long spin periods of NSs can be adequately reproduced by magnetar $+$ fallback disk model.
	In previous simulations of the long-period pulsars \citep[e.g.][]{Gencali2022,Gencali2023}  with the fallback disk model, it could reproduce the source properties at ages about $6 \times 10^5$ yr with the sources active.
	It's consistent with our results since the maximum lifetime of active disk is about $1.3 \times 10^6 {\rm ~yr}$ in our simulation, while its proportion is extremely low. 
	On the other hand, the long-period pulsars are extremely rare and peculiar, so they obviously require longer-active disks.
	
	As is shown in Figure \ref{fig:hist}, the active lifetime of fallback disk is very short.
	In addition, some newborn NSs with high magnetic fields often exhibit strong X-ray emissions due to magnetic energy release, so it's very hard for us to detect the X-ray emissions resulting from the disk.
	However, the disk in noninteracting state is thought to be cool enough which emits mainly in infrared rather than X-ray. So the James Webb Space Telescope (JWST), which is already in orbit, may provide many new surprises for the exploration of the debris disk \citep{Swan2024}.
	
	\section{Summary}
	
	In this work, we explore the lifetime of supernovae fallback disk around isolated NSs.
	We assume the disk's mass varying in a self-similar way and its interaction with the NS takes place in accretion or propeller phase while not in ejector phase or dead state.
	Our results show that most of the fallback disk can exist around the NS for a very long time and some of them can accompany through the whole life of the NS, while the interaction lifetime is much shorter, nearly all of which are less than $0.1 {\rm ~Myr}$.
	The four exceptional sources (4U 0142, 1E 1613, J1627 and J1839) lie at the edge or outside the active fallback disk regions in our findings, suggesting that their disks might be in noninteracting states.
	
	\begin{acknowledgments}
		We thanks the anonymous referee for helpful comments.
		This work was supported by the Natural Science Foundation of China under grant No. 12203051, 12273014, 12041301, 12121003, 11988101, the National Key Research and Development Program of China (2021YFA0718500) and the Natural Science Foundation (under grant No. ZR2023MA050) of Shandong Provence. 
	\end{acknowledgments}
	
	

	\appendix
	
	\section{Parameter study}
	
	\begin{figure}
		\plotone{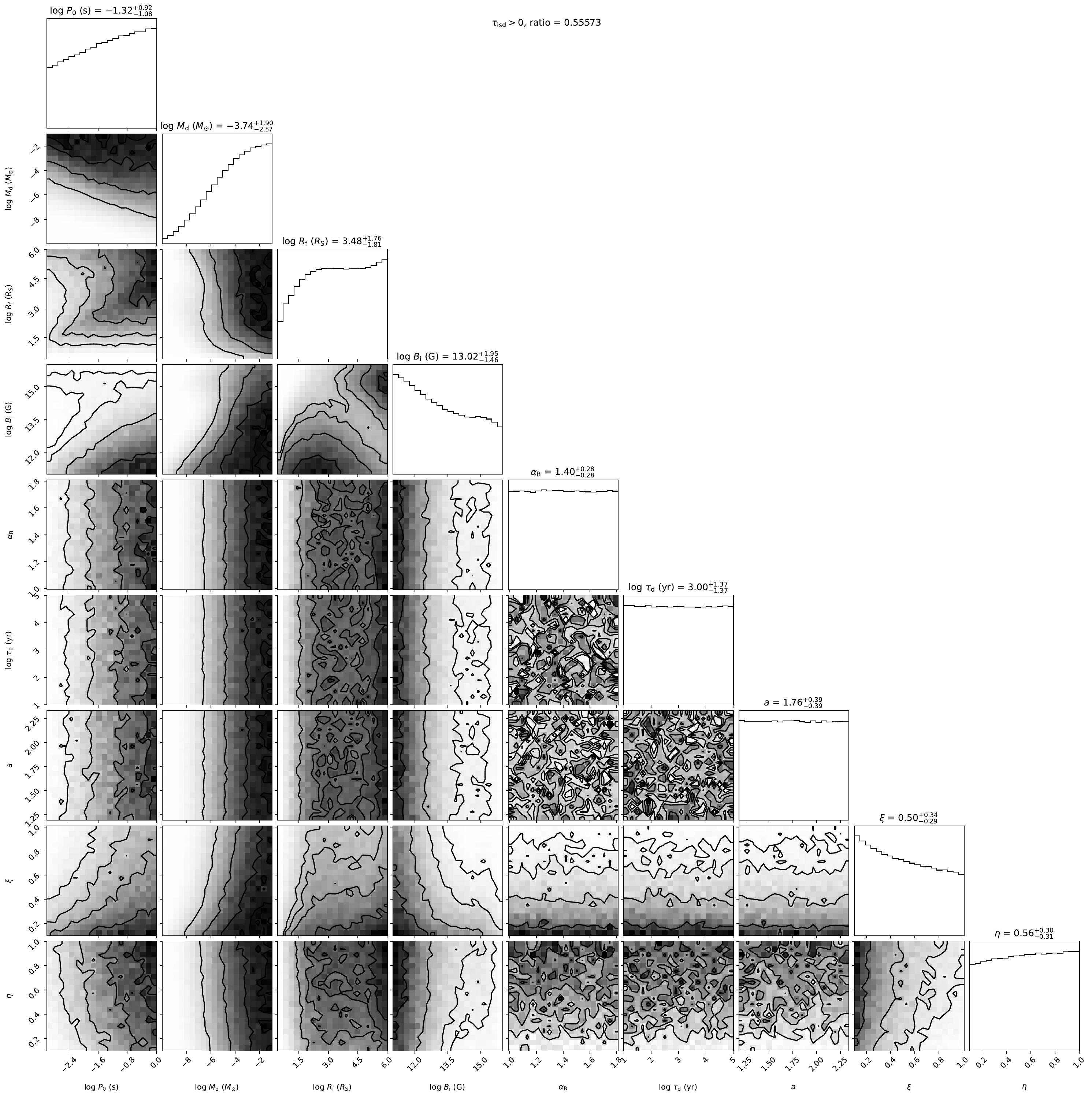}
		\caption{Parameters distribution with the interacting lifetime of fallback disks $\tau_{\rm isd}>0$.
			\label{fig:tau_9p}}
	\end{figure}
	
	Table \ref{table:param} shows the ranges of input parameters, which are all in broad ranges.
	The initial spin periods and magnetic fields of NSs are simulated to be in the range of $\sim 0.01 - 0.1 {\rm ~s}$ and $\sim 10^{12}-10^{14} {\rm ~G}$, respectively \citep[e.g.][]{Faucher-Giguere2006,Gullon2014,Igoshev2022,Xu2023,Du2024}.
	We take wide ranges of $\sim 0.001 - 1 {\rm ~s}$ and $\sim 10^{11}-10^{16} {\rm ~G}$ in this work for covering some special ones, e.g., magnetars with strong magnetic fields.
	The parameters about magnetic fields decay, which are the characteristic decay timescale $\tau_{\rm d}$ and the power law index $\alpha_{\rm B}$, are taken to be $10-10^5 {\rm ~yr}$ and $1.0-1.8$ referred to previous works \citep{Fu2012,Xu2023}.
	The parameter $\xi$, which is the correction one between the inner radius and the traditional Alfv\'en radius for spherical accretion, is usually taken to be $0.5$ \citep{Ghosh1979,Long2005} or $1.0$ \citep{Wang1996}, and we use a wide range of $0.1-1.0$.
	Finally, the four parameters about the fallback disk, i.e., the initial mass $M_{\rm d,0}$, the initial outer radius $R_{\rm f}$\footnote{The self-gravity radius $R_{\rm sg} \approx 9.62 \times 10^{10} {\rm ~cm}$ \citep{Xu2019}, so here we set the upper limit of $R_{\rm f}$ as $10^6 R_{\rm S} \approx 4 \times 10^{11} {\rm ~cm}$. }, the mass loss rate $\eta$ and the power law index of mass loss rate $a$, are relatively less studied.
	Especially, there are some debates on patametar $a$ \citep{Xu2021}, which depends on wind loss and opacity of the disk.
	In general, $a$ is taken to be $4/3$ in the case without disk wind and $4/3 < a \leq 5/3$ on the contrary \citep[e.g.][]{Pringle1991,Cannizzo2009,Liu2015,Lin2021}. However \cite{Beniamini2020} suggests that $a>5/3$ for radiatively inefficient accretion flows.
	On the other hand, $a = 19/16$ and $1.25$ are thought to be opacity dominated by electron scattering and Kramers opacity \citep[e.g.][]{Cannizzo1990,Li2007,Tong2016}, respectively. 
	So we take a broad range of $19/16-7/3$ for the parameter $a$ \citep{Xu2021}.
	The other three parameters about the fallback disk are also taken in broad ranges \citep{Xu2019}.

	\begin{figure}
		\plotone{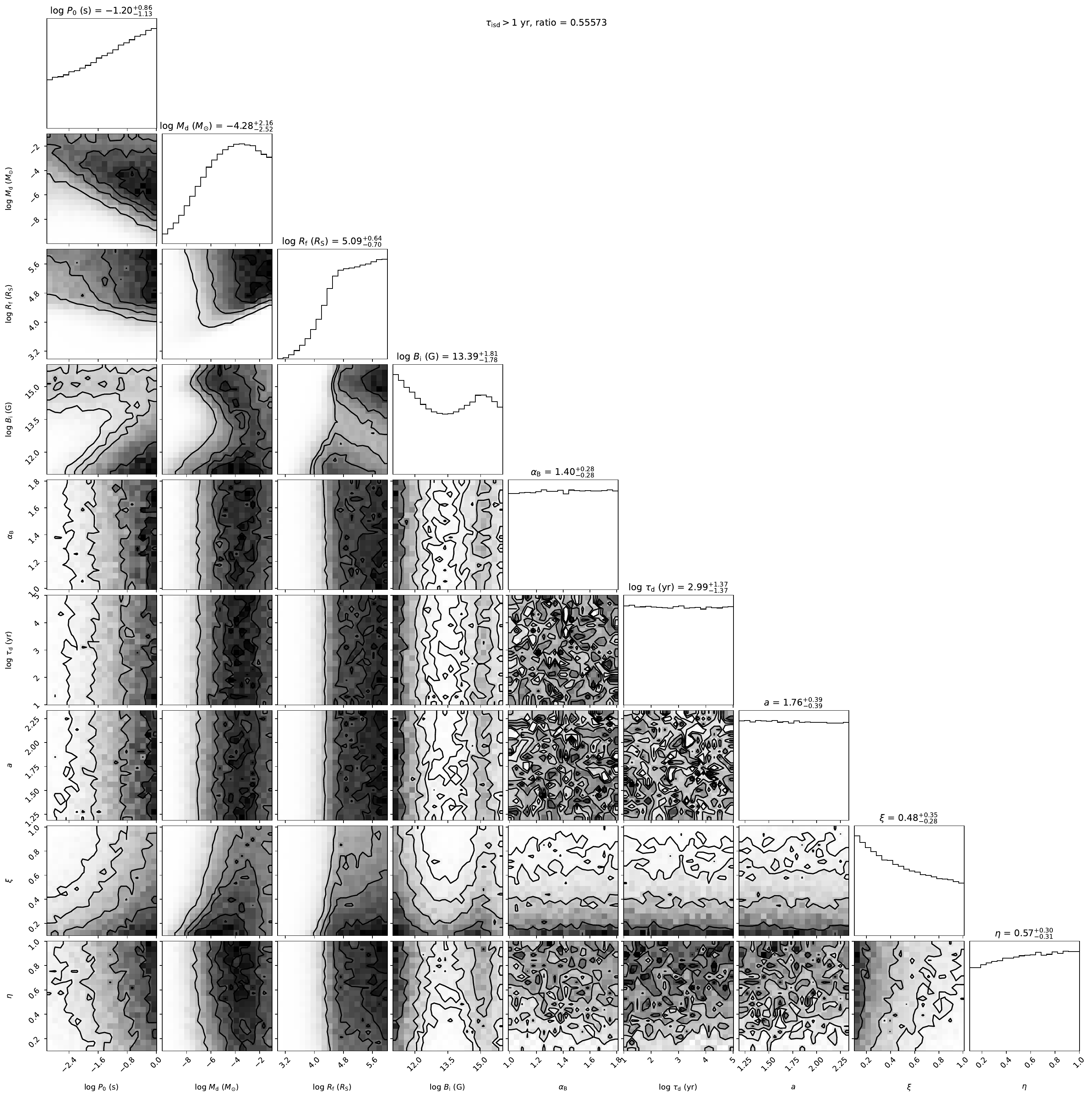}
		\caption{Same as Figure \ref{fig:tau_9p} but with $\tau_{\rm isd}>1 {\rm ~yr}$.
			\label{fig:tau_1yr}}
	\end{figure}
	
	\begin{figure}
		\plotone{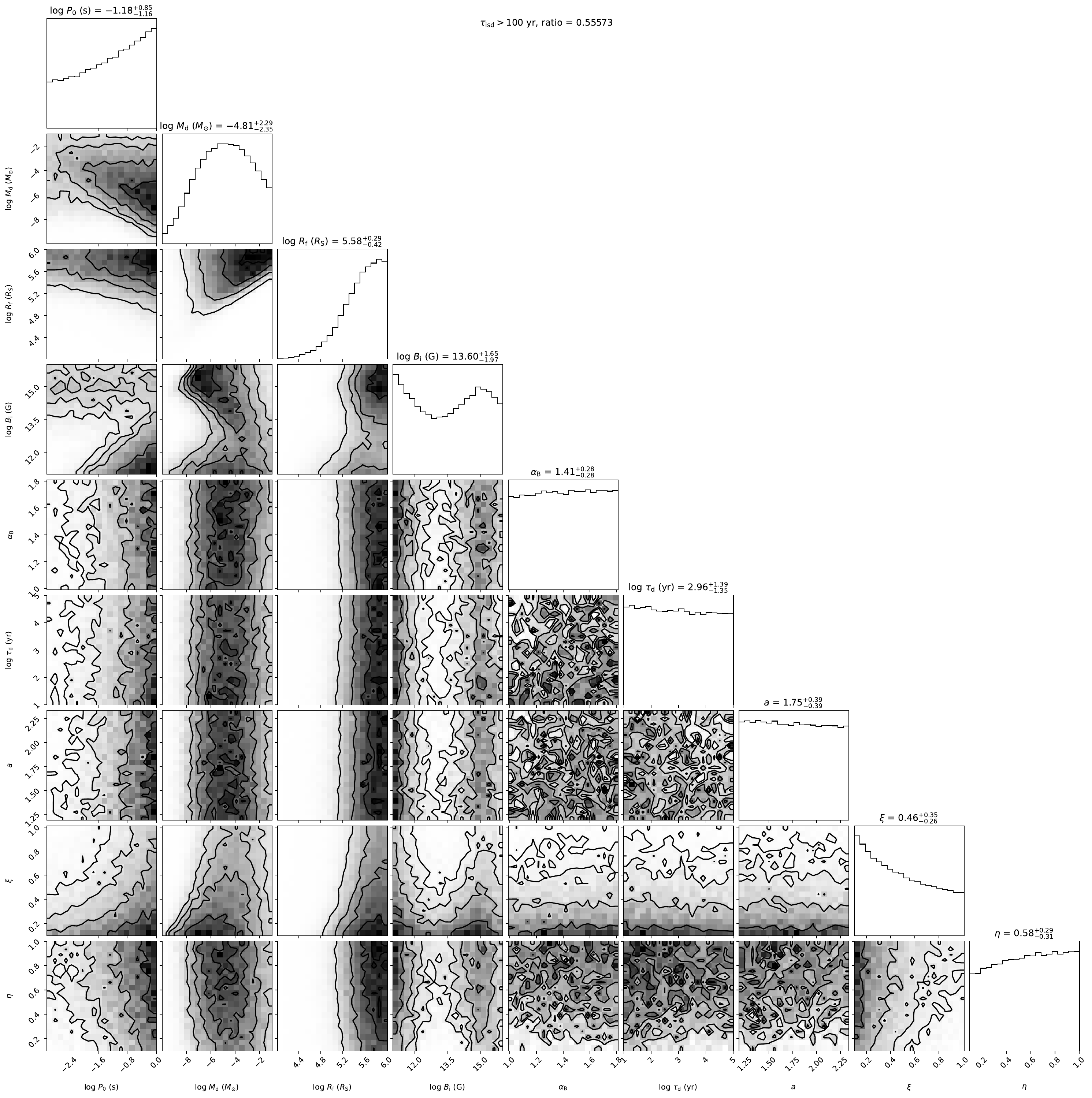}
		\caption{Same as Figure \ref{fig:tau_9p} but with $\tau_{\rm isd}>100 {\rm ~yr}$.
			\label{fig:tau_100yr}}
	\end{figure}

	We show the parameters distribution with the interacting lifetime of fallback disks $\tau_{\rm isd}>0$, $1 {\rm yr}$ and $100 {\rm yr}$ in Figure \ref{fig:tau_9p}, \ref{fig:tau_1yr} and \ref{fig:tau_100yr}, respectively.
	It shows that 
	\begin{enumerate}
		\item The distribution of $\alpha_{\rm B}$, $\tau_{\rm d}$, $a$ and $\eta$ are nearly uniform, which means that these parameters have minor impact on the interacting lifetime of fallback disks.
		Intuitively, the parameters $a$ and $\eta$ were thought to have significant impact on the results. In fact, they have great influence on the existence lifetime of the fallback disk while affect little on the interacting lifetime because it's much shorter than the former one.
		\item The distribution of $P_0$ and $R_{\rm f}$ are increasing as their value increases, which means that the interacting disks prefer larger $P_0$ and $R_{\rm f}$.
		It's apparent because the fallback disk is more likely to form for a slow rotation NS and large initial disk radius of the disk, and stay in active state for a longer time learning from the 3rd condition for interacting/active disk.
		\item On the contrary, the distribution of $\xi$ is reducing as its value increases, which means that the interacting disks prefer smaller $\xi$.
		The smaller $\xi$ means smaller inner radius of the disk, so it's similar to the previous one that the fallback disk can remain in interacting state for an extended period, as demonstrated by the third condition for interacting disks.
		\item The distribution of $M_{\rm d}$ is increasing as its value increases for all the interacting disks (Figure \ref{fig:tau_9p}) while has a peak at $\sim 10^{-4.28}$ and $10^{-4.81}$ ${\rm M_{\odot}}$ for $\tau_{\rm isd}>1 {\rm yr}$ (Figure \ref{fig:tau_1yr}) and $100 {\rm yr}$ (Figure \ref{fig:tau_100yr}). This tendency is consistent with the  $M_{\rm d}$ distributions in \citet{Xu2019} and \citet{Xu2021}, which is because that the NS stays in ejector phase too long with a slighter fallback disk and experiences too long accretion with a more massive disk making a rapid mass loss.
		\item The distribution of $B_{\rm i}$ is the most extraordinary. It reduces as its value increases first, then has a peak at $\sim 10^{15} {\rm ~G}$. The reason may be that the weak magnetic fields are friendly for disk formed while the strong fields make disk existing longer.
	\end{enumerate}
	
	\bibliography{sample631}{}
	\bibliographystyle{aasjournal}

\end{document}